\documentclass[twocolumn,showpacs,preprintnumbers,amsmath,amssymb,widetext,super
scriptaddress]{revtex4}
\usepackage{graphicx}
\begin{document}
\title{Ultra-low dissipation Josephson transistor}
%\title{Controlling supercurrents through cold electrons}
\author{Francesco Giazotto}
\email{giazotto@sns.it}
\affiliation{NEST-INFM \& Scuola Normale
Superiore, I-56100 Pisa, Italy}
\author{Fabio Taddei}
\email{taddei@sns.it}
\affiliation{NEST-INFM \& Scuola Normale
Superiore, I-56100 Pisa, Italy} \affiliation{ISI Foundation, Viale
Settimio Severo, 65, I-10133 Torino, Italy}
\author{Tero T. Heikkil\"{a}}
\affiliation{Low Temperature Laboratory, Helsinki University of Technology, P.O.
Box 2200, FIN-02015 HUT, Finland}
\author{Rosario Fazio}
\author{Fabio Beltram}
\affiliation{NEST-INFM \& Scuola Normale Superiore, I-56100 Pisa, Italy}

\begin{abstract}
A superconductor-normal metal-superconductor (SNS) transistor based on
superconducting microcoolers is presented. The proposed  4-terminal device
consists of a long SNS Josephson junction whose N region is in addition
symmetrically connected to superconducting reservoirs through tunnel barriers
(I). Biasing the SINIS line allows to modify the quasiparticle temperature in
the weak link, thus controlling the  Josephson current. We show that, in
suitable voltage and temperature regimes, large supercurrent enhancements can be
achieved  with respect to equilibrium, due to electron ``cooling'' generated by
the control voltage. The extremely low power dissipation intrinsic to the
structure  makes this device relevant for a number of electronic applications.
\end{abstract}


\maketitle

In the last few years thermoelectric cooling has gained an increased attention
in solid-state physics both from the basic research point of view and as a tool
for cryogenic applications, in particular for cooling space-based infrared
detectors \cite{cryo,luukanen}. The advantage of electronic microrefrigerators
stems from their intrinsic compactness, little power dissipation and ease of on-chip integration.
To date several refrigeration schemes at cryogenic
temperatures have been proposed \cite{QD,nahum,FS,paterson,kap}. In particular,
cooling schemes based on superconductor-insulator-normal metal (SIN) tunnel
junctions proved  extremely effective in refrigerating both electrons
\cite{nahum,leivo} and phonons \cite{manninen}. In such systems the cooling
principle is based on the superconducting energy gap ($\Delta$) that allows to
remove quasiparticles with higher energies from the normal electrode more
effectively than those with lower energy \cite{averin}.

\begin{figure}[h!]
\begin{center}
\includegraphics[width=7cm]{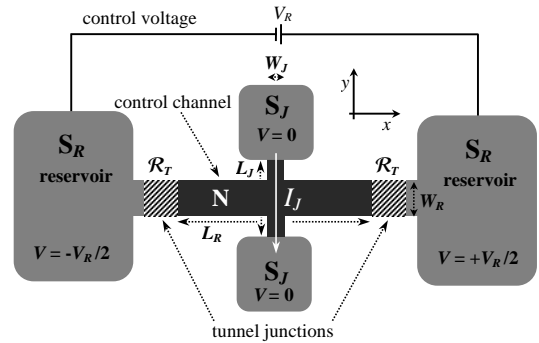}
\end{center}
\caption{Sketch of the proposed SNS transistor. The S$_J$N junctions are assumed
to be perfectly transmissive electric contacts while the superconducting
reservoirs (S$_R$) are connected to the N region through identical tunnel
barriers of resistance $\mathcal R_T$. The bias $V_R$ allows to tune the
supercurrent $I_J$ in the weak link, both increasing and decreasing its
magnitude with respect to equilibrium.}
\label{devicescheme}
\end{figure}

In this letter we propose a device that exploits thermoelectric cooling to
obtain transistor effect by means of a symmetric SIN double-microcooler
integrated with a SNS weak link (see Fig.\ref{devicescheme}).
In existing \emph{controllable} Josephson junctions critical supercurrent
modulation is achieved by driving the quasiparticle distribution out of
equilibrium \cite{vanWees,wilhelm,yip} by dissipative current injection into the
N region of the junction \cite{morpurgo, baselmans,shaik,birge}. This leads to
controlled supercurrent suppression.
Our scheme exploits the exponential temperature dependence of the critical
supercurrent in long SNS weak links \cite{zaikin,wilhelm2}. We show that thanks
to electric cooling, critical supercurrent values can be strongly increased from their equilibrium magnitude
yielding a substantial current gain in the final device. Importantly, thanks to
its unique working principle our device leads to control currents that are
orders of magnitude lower than in all-normal control lines. This property makes
this device of particular interest for extremely-low dissipation cryogenic
applications.

The 4-terminal device (sketched in Fig.~\ref{devicescheme}) consists of a N
region laterally  connected (in the $y$-direction) to two superconductors
(S$_J$) through perfectly transmitting interfaces \cite{interface}, thus
realizing a SNS junction of length $L_J$ and width $W_J$.
We consider a mesoscopic diffusive SNS weak link (i.e., $\ell _{m} < L_{J} <
\ell _{\varphi}$, where $\ell_{m}$ and  $\ell_{\varphi}$ are the elastic mean
free path and the single-particle dephasing length, respectively) and, for
reasons that will be clear in the following, we assume that the  junction is
\emph{long} (i.e., $\Delta_{J} \gg \hbar \mathcal{D}/ L_{J}^{2}=E_{Th}$, where
$\Delta_{J}$ is the S$_J$ energy gap, $\mathcal{D}$ is the N region diffusion
constant and $E_{Th}$ is the Thouless energy).
Along the $x$-direction the N region is connected to two additional
superconducting reservoirs (S$_R$) through identical tunnel barriers (I) of
resistance $\mathcal R_T$, thus realizing a SINIS symmetric double-refrigerator
(cooling control line) \cite{leivo,nahum} of length $L_R$ and width $W_R$.
Resistance
$\mathcal R_T$ is taken to be much larger than the N region resistance in the
$x$-direction so that no voltage drop will occur in the N region, thus
optimizing the cooling power of the SINIS device \cite{averin}.
We also assume, following Ref. \cite{tero}, that $W_R \ll L_J$ and $W_J \ll
L_R$.
 Moreover, we suppose that  S$_J$  are kept at the same potential, chosen to be
zero, while  S$_R$  are kept at equal and opposite voltages $V=\pm V_{R}/2$.

We are interested in the regime where the quasiparticle distribution retains a
local thermal equilibrium manifesting itself as a Fermi-Dirac function with
effective electron temperature $T_e$ differing in general from the lattice
temperature $T_l$ \cite{baselmansthesis}. This regime can be attained by
assuming a large enough $L_R$ (i.e., $\ell_{e-e}< L_R < \ell_{e-ph}$, where
$\ell_{e-e}$ and $\ell_{e-ph}$ are the electron-electron and electron-phonon
scattering lengths, respectively) so that injected quasiparticles may thermalize
through electron-electron interaction. For this to be true we need to focus on
lattice temperatures $T_l < 1$ K, where electron-electron scattering is the main
relaxation mechanism.
For $L_R \gg \ell_{\varphi}, \ell_{e-e}$, the cooling control line can be
considered as composed by two SIN junctions incoherently added in series.
Under bias voltage $V_R $ the power $\mathcal{P}_N$ transferred  from the N
region to the superconductors S$_R$ can be calculated along the lines of Ref.
\cite{averin}.
The main results of this analysis are that $\mathcal{P}_N$ is a symmetric
function of $V_R$, with a maximum slightly below $V_R =\pm 2\,\Delta_R/e $. As a
function of the lattice temperature, $\mathcal{P}_N$ shows a maximum of
$0.12\,\Delta_{R}^2/e^2 \mathcal{R}_T$ at $T_l \approx 0.3\,\Delta_R /k_B$,
quickly decreasing at both lower and higher temperatures.
As a consequence, both small $\Delta_R$ and low $\mathcal R_T$ are desirable in
order to achieve high cooling power at sub-Kelvin temperatures.
The final $T_e$ in the weak link is determined by considering those mechanisms
that can transfer energy into the N region. In our case the main contribution is
related to electron-phonon scattering given by $\mathcal P_{e-l}=\Sigma \mathcal
V(T_e^5-T_l^5)$ \cite{urbina}, where $\mathcal V$ is the N region volume and
$\Sigma$ is a material-dependent parameter. The effective temperature $T_e
(V_R)$ is then determined solving the energy-balance equation $\mathcal
P_N(V_R,T_e,T_l)+\mathcal P_{e-l}(T_e,T_l)=0$. In writing the above expression
we set the quasiparticle temperature in the superconducting electrodes equal to
$T_l$. Although this is an idealized assumption we expect it, however, to be a
reasonable approximation in real devices exploiting quasiparticle traps
\cite{anghel}.

The supercurrent in the diffusive SNS junction ($I_J$) can be calculated from
the equation  \cite{yip,wilhelm,tero}:
\begin{equation}
I_J=\frac{\sigma_N\mathcal A}{eL_J}\int^{\infty}_{0}f_A(\varepsilon,T_e)\,j_S
(\varepsilon,\phi,T_l)\,d \varepsilon,
\label{supercurrent1}
\end{equation}
where $\sigma_N$ is the  conductivity of the weak link, $\mathcal A$ its cross
section and $\phi$ is  the phase difference between the superconductors.
According to  (\ref{supercurrent1}) $I_J$ is expressed in terms of the spectral
supercurrent $j_S (\varepsilon,\phi,T_l)$ weighted by the antisymmetric part of
the quasiparticle distribution function in the weak link $f_A(\varepsilon,T_e)$
(note that the symmetric part does not contribute since $j_S$ is an odd function
of energy).
In the present case of quasi-equilibrium  the latter reduces to
$f_A(\varepsilon,T_e)=\tanh[\varepsilon/2k_BT_e]$.
It is worth noticing that  $j_S$ depends on the lattice temperature $T_l$
through $\Delta _J$, while $f_A$ depends on the electron effective temperature
$T_e$ in the weak link.
The working principle of our controllable junction is described in a simple way
in Fig. 2(a, b), where both $j_S$ (dash-dotted line) and $f_A$ (solid line) are
plotted on the same graph as a function of normalized energy.
Hatched areas represent the integral of their product, composed of a positive
and a negative contribution.
In Fig. 2(a) $T_e$ is set equal to $T_l$, obtained with $V_R =0$, while in Fig.
2(b) $T_e$ is lowered below $T_l$ by biasing the control line, therefore
sharpening the quasiparticle distribution. As a result,  the supercurrent is
enhanced since the positive area is increased \cite{expression}. This is
demonstrated in the following by numerically solving the energy-balance equation
for $T_e(V_R)$ and inserting it in the expression for $I_J(V_R)$ in
(\ref{supercurrent1}).
\begin{figure}[h!]
\begin{center}
\includegraphics[width=8cm]{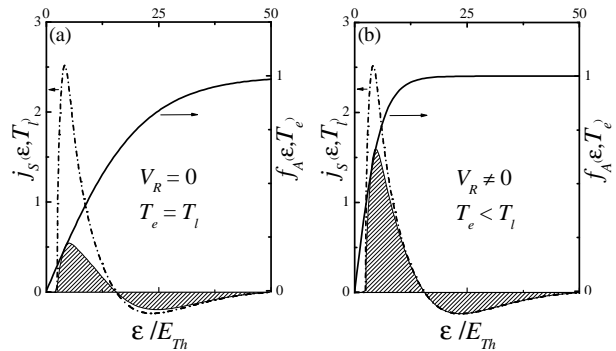}
\end{center}
\caption{Qualitative behavior of both the supercurrent spectral density ($j_S
(\varepsilon ,T_l)$) and the quasiparticle distribution function
($f_A(\varepsilon ,T_e)$). (a) For $V_R =0$ (i.e., in equilibrium) the electron
temperature ($T_e$) equals that of the lattice ($T_l$). Shaded area represents
the contribution to supercurrent originating in such a configuration. (b) Upon
biasing ($V_R \neq 0$), the \emph{cooling} effect of the SINIS line lowers $T_e$
with respect to $T_l$ allowing $f_A(\varepsilon ,T_e)$ to become sharper. The
result is a supercurrent enhancement (see text).}
\label{ARscheme}
\end{figure}

In order to simulate a realistic device we choose Al for the S$_R$ reservoirs
($\Delta_R=180$ $\mu$eV ), Cu for the N region, $L_R=10$ $\mu$m, $W_R =0.5$
$\mu$m, $L_J=1$ $\mu$m and $W_J =0.4$ $\mu$m. Supposing a 100-nm-thick film, the
total volume of N region amounts to $\mathcal V=0.52$ $\mu$m$^3$. For metals
like Ag, Cu, and Au the chosen value for $L_R$ corresponds to the interacting
hot-electron regime \cite{morpurgo}.
Furthermore we take $\Sigma=2\times 10^{-9}$ W K$^{-5}$ $\mu$m$^{-3}$
\cite{nahum} and $\mathcal R_T=100$ $\Omega$ \cite{blamire}. For S$_J$  we
choose Nb  ($\Delta_J=1.52$ meV) and assuming a diffusion constant $\mathcal
D=0.007$ m$^2$/s  \cite{pierre}, the Thouless energy is $E_{Th}\simeq4.6$
$\mu$eV, so that
the considered junction is in the long limit ($\Delta_J/E_{Th}\simeq 330$)
\cite{nbnote}.
\begin{figure}[h!]
\begin{center}
\includegraphics[width=8cm]{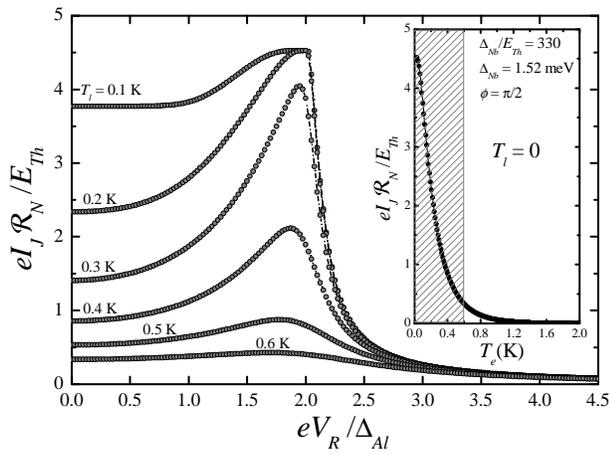}
\end{center}
\caption{Normalized critical current vs voltage $V_R$ of the Nb/Cu/Nb weak link
calculated for several lattice temperatures $T_l$. In this configuration a
supercurrent enhancement of about 300$\%$ can be achieved at $T_l =300$ mK due
to quasiparticle cooling. For  $V_R >2\,\Delta _R /e$, efficient supercurrent
suppression occurs due to electron heating. The inset shows the junction
supercurrent vs temperature characteristic calculated at $\phi =\pi /2$. Hatched
region shows a temperature range suitable for efficiently tuning the transistor.}
\label{critcurvsv}
\end{figure}
In Fig.~\ref{critcurvsv} we plot the calculated normalized  $I_J(V_R)$
characteristic for several starting lattice temperatures ($\mathcal
R_N=L_J/\sigma_N \mathcal A$ is the SNS junction normal state resistance).
For all chosen $T_l$, $I_J(V_R)$ increases monotonically for bias up to $eV_R
\simeq 2\Delta_{Al}$, for which the effect of the control line is to decrease
the effective electron temperature. In the case of an \emph{all-normal} control
line (i.e., without S$_R$) the effect would have been only to enhance the
electron temperature ($T_e > T_l$) \cite{morpurgo} with consequent suppression
of the supercurrent.
For larger bias, in fact, an efficient suppression of the supercurrent occurs
due to electron heating. As a result the device behaves as a fully-tunable
superconducting transistor.
At zero temperature a critical current of about $54\,\mu$A can be achieved (it
can be further enhanced in a slightly shorter weak link, resulting into a larger
$E_T$ and a smaller $\mathcal R_N$ \cite{tero}).
Remarkably, the current gain $\mathcal{G}=\delta I_J/I_{R\,max}$, i.e., the
ratio between the maximum supercurrent enhancement at a given $T_l$ and the
injected quasiparticle current needed to produce such enhancement, obtains
values as large as $\mathcal G \simeq 90$ in the examined $T_l$ range.
Note that $\mathcal G$ is  dependent on the choice of the structure parameters.

We can now clarify the reason why \emph{long} junctions are more appropriate for
the device to work.
Since in long junctions  \cite{tero} $E_{Th}$ sets the temperature scale for the
supercurrent, for sufficiently low lattice temperatures  $I_J$ is independent of
$T_l$ and  only depends on $T_e$ as displayed in the inset of Fig. 3:
by starting from a given temperature $T_e$ at $V_R = 0$, an increase of $V_R$
lowers $T_e$ thus enhancing $I_J$. This shows  that it is advisable to choose a
temperature window where the slope of the $I_J(T_e)$ characteristic is large in
the range of a high cooling power from the control line, in order to achieve
large supercurrent enhancements (of the order of several hundreds percent) with
small temperature variations. This optimized temperature range  is marked by the
hatched area and is well within the cooling performance of the chosen
refrigerators. Note that in the \emph{short} junction limit ($\Delta_J \ll
E_{Th}$), it is the energy gap $\Delta_J(T)$ which sets the temperature
dependence of the supercurrent, thus implying a  much reduced effect from the
cooling line.

We would like, furthermore, to comment on the impact of power dissipation in
this device for practical applications. For fixed geometry a straightforward
comparison shows that, in the present case, control currents $I_R$ corresponding
to bias larger than $2\Delta _R/e$ are at least two orders of magnitude lower
($I_R \approx 10^{-6}$ A) \cite{corrente} than in the existing \emph{all-normal}
controllable Josephson junctions \cite{morpurgo, baselmans,shaik,birge}.
For bias corresponding to supercurrent enhancement, on the other side,   control
currents are at least four orders of magnitude lower, due to the  dependence of
tunneling subgap current from barrier resistance. These considerations suggest
that such a device can be well suited for  high-density and ultra-low
dissipation cryogenic applications.

In conclusion, a promising scheme to obtain transistor effect in a Josephson weak link
was presented. A realistic Nb/Cu/Nb weak link in combination with
Al/Al$_2$O$_3$/Cu refrigerators was analyzed and we demonstrated that a
supercurrent enhancement of several hundreds percent can be achieved.
Moreover, the drastic reduction of control currents intrinsic to the proposed
design with respect to existing schemes makes such a structure of interest for
applications where extremely low power dissipation is required at cryogenic
temperatures.

The authors gratefully acknowledge J. P. Pekola for fruitful discussions.
The present work was supported in part by INFM under the PAIS project TIN.

%------------------------------------------ References

%\begin{references}

%\end{references}

%------ Figures beginning -------------

%\begin{figure}
%\caption{}
%\label{F1}
%\end{figure}

%\begin{figure}
%\caption{
%}
%\label{F2}
%\end{figure}

%\begin{figure}
%\caption{
%}
%\label{F3}
%\end{figure}


\begin{thebibliography}{25}
\bibitem{expression}
    Note that for $k_B T_e \gg E_{Th}$, $I_J=I_C \sin(\phi)$ with
$I_C=\frac{64 \pi}{3+2\sqrt{2}}\frac{T_e}{e\mathcal{R}_N}
\sqrt{\frac{2\pi T_e}{E_{Th}}} \exp\left(-\sqrt{\frac{2\pi
T_e}{E_{Th}}}\right)$ \cite{zaikin,wilhelm2}. Hence, in this limit
the critical current of the junction depends exponentially on the
effective electron temperature.
\bibitem{nbnote}
        For the SNS junction it is important to choose a superconductor with a
gap larger than the operating voltage of the SINIS cooler, thus ensuring thermal
isolation of the S$_J$ reservoirs from the weak-link region to be cooled.
\bibitem{corrente}
    This can be further reduced by choosing appropriate higher tunnel barrier
resistances.
\end{thebibliography}
\end{document}